\documentclass[aps, prl, reprint, graphics, showpacs, amssymb, amsmath, superscriptaddress]{revtex4-1}
\usepackage{mathrsfs}
\usepackage{bm}
\usepackage{times}
\usepackage{graphicx}
\usepackage[colorlinks,citecolor=blue,linkcolor=blue]{hyperref}
\usepackage{array}
\usepackage{float}
\usepackage{colortbl}
\usepackage{multirow}
\usepackage{tabularx}
\usepackage{ulem}

\begin{document}

\title{Topological Electride Phase of Sodium at High Pressures and Temperatures}

\author{Busheng Wang}\affiliation{Department of Chemistry, State University of New York at Buffalo, Buffalo, NY 14260-3000, USA}
\author{Katerina P. Hilleke}\affiliation{Department of Chemistry, State University of New York at Buffalo, Buffalo, NY 14260-3000, USA}
\author{Xiaoyu Wang}\affiliation{Department of Chemistry, State University of New York at Buffalo, Buffalo, NY 14260-3000, USA}
\author{Danae N. Polsin}\affiliation{University of Rochester Laboratory for Laser Energetics, Rochester, New York 14623, USA}\affiliation{Department of Mechanical Engineering, University of Rochester, Rochester, New York 14627, USA}
\author{Eva Zurek}\email{ezurek@buffalo.edu}\affiliation{Department of Chemistry, State University of New York at Buffalo, Buffalo, NY 14260-3000, USA}

\date{\today}

\begin{abstract}
\textit{Ab initio} evolutionary structure searches coupled with quasiharmonic calculations predict that the insulating Na $hP$4 phase transitions to a novel $\mathit{P}6_{3}/\mathit{m}$ phase between 200~GPa at 150~K, and 350~GPa at 1900~K. $\mathit{P}6_{3}/\mathit{m}$ Na is a topological semimetal with a Dirac nodal surface that is protected by a non-symmorphic symmetry, $S_{2z}$. It is characterized by localized non-nuclear charge within 1D honeycomb channels and 0D cages rendering it an electride. These results highlight the complexity of warm dense sodium’s electronic structure and free energy landscape that emerges at conditions where ionic cores overlap.
\end{abstract}
\maketitle

%\section*{Introduction}%The counterintuitive high-pressure behavior of the alkali metals continues to fascinate and surprise \cite{rousseau2011exotic}. %
Since the dawn of quantum mechanics, it has been assumed that at sufficiently high pressures all matter would adopt simple structures with densely packed ionic cores and nearly free electrons. However, unprecedented structural complexity and anomalous physical properties have been theoretically predicted and observed in the phase diagrams of alkali metals \cite{rousseau2011exotic}. In sodium, unexpected behavior is found already at ambient pressure where the $bcc$ structure undergoes a partial martensitic phase transition upon cooling \cite{elatresh2020fermi}.  Compression to $\sim$120~GPa results in an astounding $\sim$700~K decrease in the melting point  \cite{gregoryanz2005melting}, whose minimum is associated with rich structural diversity \cite{Gregoryanz:2008a,mcmahon2007structure}. Near 200~GPa and room temperature an insulating $hP$4 phase  has been made, \cite{ma2009transparent} persisting upon ramp-compression to nearly 500~GPa \cite{Denae2022Na}.

This extraordinary behavior of Na was foreseen by Neaton and Ashcroft who postulated that at pressures large enough to induce 2$p$ orbital overlap, sodium's valence electrons would be impelled into the interstitial regions of its crystal lattice due to orthogonality and Pauli repulsion, resulting in a metal to insulator transition \cite{neaton2001constitution}. About a decade later this prediction was verified in static compression experiments that reported a double hexagonal $hP$4 structure ($\mathit{P}6_{3}/\mathit{mmc}$, No.\ 194) with a band-gap exceeding 1.3~eV \cite{ma2009transparent}. The $hP$4 phase is characterized by electrons that are strongly localized at lattice voids -- rendering it a prototypical example of a high-pressure electride, where the ionic cores play the role of cations, and interstitial electrons act as anions. Such interstitial electron localization upon densification has been explained by core exclusion and proximity \cite{rousseau2008interstitial}, $p$-$d$ hybridization \cite{ma2009transparent}, and by comparing the pressure dependence of orbitals centered on interstitial quasiatoms to those centered on ionic cores \cite{miao2014high,Zurek:2019k}. Density functional theory (DFT) calculations have shown that sodium's unusual optical properties \cite{gatti2010sodium},  melting behavior, and rich polymorphism are consequences of the electride state \cite{marques2011optical}. First-principles molecular dynamics simulations find that localized electron bubbles persist in the fluid at high temperatures and pressures \cite{paul2020thermal,Bonev:2007a}. 

Moving to denser structures yet, cold DFT calculations predict that Na will transition from an insulating $oP$8 phase to a metallic $cI$24 phase characterized by Na$_{12}$ icosahedra near 15.5~TPa \cite{li2015metallic}.  Comparison of the Gibbs free energies of these phases obtained within the quasiharmonic approximation at 20~TPa and 1000~K led to the conclusion that the effect of the temperature on the phase transition pressure is negligible. 

In contrast, laser-driven ramp-compression experiments of Na to nearly 500~GPa and $\sim$3000~K hint of temperature-dependent structural complexity at these conditions \cite{Denae2022Na}. In this experiment the thermodynamic pathway led to pressure-driven melting and recrystallization on nanosecond timescales. \textit{In situ} X-ray diffraction (XRD) revealed a series of phases upon recrystallization, with peaks attributed to Na $hP$4 evident only above 400~GPa. At lower pressures ($\sim$240-325~GPa) the obtained diffraction peaks could not be explained by the $hP$4 structure.  Instead, they were attributed to a $cI$16 phase, isostructural with known phases of Li and Na found at lower temperatures and pressures, and a suggested $R\bar{3}m$ symmetry structure. However, our DFT calculations (see the Supplemental Material \cite{SM3}), showed that both structures are dynamically and thermally unstable at these conditions. Therefore, further studies are essential to uncover the unknown Na phases created in Ref.\  \cite{Denae2022Na} during ramp-compression of Na at intermediate pressures.

Theory has suggested that compressed Group I elements and alloys may possess topological properties. For example, DFT calculations concluded that forms of dense hydrogen \cite{naumov2016} and Na $hP$4 \cite{naumov2017metallic}  both possess metallic surface states,  the band structure of Li$_5$H presents Dirac-like features  \cite{Zurek:2012a}, and various phases of Li at intermediate pressures were computed to be topological semimetals with nodal loops or lines in the vicinity of the Fermi level, $E_F$ \cite{mack2019emergence,elatresh2019high}. The interest in topological semimetals stems from their unique properties including high mobilities \cite{novoselov2004electric}, giant magnetoresistance \cite{liang2015ultrahigh}, and because their massless fermions make it possible to simulate intriguing high-energy and relativistic physics phenomena in table-top experiments \cite{guan2017artificial}. Topological semimetals may be characterized by the presence of 0-dimensional (0D) point nodes, 1-dimensional (1D) nodal lines, or 2-dimensional (2D) nodal surfaces where the conduction and valence bands cross in the Brillouin Zone (BZ) \cite{lv2021experimental}.

This Letter presents an investigation into the structural complexity, topological features and core-electron chemistry emerging in warm dense sodium. We unravel the crystal structure of the unknown Na phase whose spectral signatures were observed during ramp-compression -- a  topological semimetal exhibiting a Dirac nodal surface and localized non-nuclear electron density. Using evolutionary algorithm searches coupled with first principles calculations, we predict several structures whose free energies fall below that of Na $hP$4 above 200~GPa at finite temperatures, consistent with recent ramp-compression experiments. A $\mathit{P}6_{3}/\mathit{m}$ Na phase is demonstrated to possess the lowest free energy at conditions typical of those accessed in laser-driven experiments, specifically between 250~GPa at 710~K, and 350~GPa at 1900~K (Figure S7). Electronic structure calculations show that its electron density is localized within honeycomb channels and at interstitial sites, rendering it a metallic electride.  The Dirac nodal surface within $\mathit{P}6_{3}/\mathit{m}$  is topologically protected by the non-symmorphic symmetry $S_{2z}$.

\begin{figure}[]
\centerline{\includegraphics[width=0.9\columnwidth]{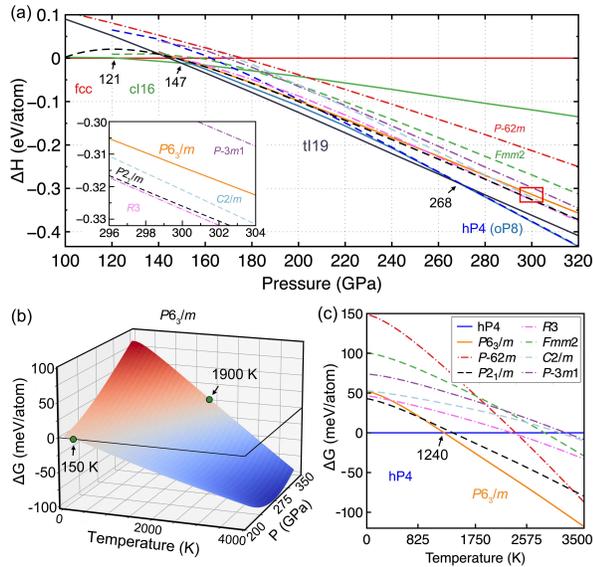}}
\caption{(Color online) (a) Enthalpies of various Na phases relative to $fcc$ ($\Delta H$). The  $fcc$ $\to$ $cI$16 $\to$ $tI$19 $\to$ $hP$4 ($oP$8) transitions are calculated at 121, 147, and 268 GPa, respectively. Besides the experimentally observed Na structures ($fcc$ \cite{hanfland2002sodium}, $cI$16 \cite{gregoryanz2008structural,mcmahon2007structure}, $oP$8 \cite{gregoryanz2008structural}, $tI$19 \cite{gregoryanz2008structural}, and $hP$4 \cite{ma2009transparent}), seven novel phases found in the evolutionary searches (five shown in the inset) were chosen for further analysis. (b) The Gibbs free energy of $\mathit{P}6_{3}/\mathit{m}$ Na relative to $hP$4 ($\Delta G$) as a function of temperature ($T$) and pressure ($P$). A negative $\Delta G$ (blue region) indicates that $\mathit{P}6_{3}/\mathit{m}$ is more stable than $hP$4. The green dots with black arrows highlight the critical conditions where $\mathit{P}6_{3}/\mathit{m}$ becomes preferred at low (150~K and 200~GPa) and high (1900~K and 350~GPa) temperatures.  (c) $\Delta G$ of various Na phases relative to $hP$4 at 300~GPa. $\mathit{P}6_{3}/\mathit{m}$  becomes the most stable phase above 1240~K.}
\label{fig:latt}
\end{figure}

To uncover the phases that were observed in experiment, we performed extensive evolutionary structure searches using the \textsc{XtalOpt} \cite{lonie2011xtalopt, falls2020xtalopt} algorithm coupled with DFT calculations \cite{SM3}. Besides the experimentally determined Na structures, seven additional dynamically stable (Section S5) low enthalpy phases found in the evolutionary runs, with $\mathit{R}3$, $\mathit{P}2_{1}/\mathit{m}$, $\mathit{C}2/\mathit{m}$, $\mathit{P}6_{3}/\mathit{m}$, $\mathit{P}\bar3\mathit{m}1$, $\mathit{Fmm}2$, and $\mathit{P}\bar62\mathit{m}$ symmetries (Table S2), were singled out for further analysis. Figure \ref{fig:latt}(a) illustrates that our calculated cold phase transition sequences and corresponding pressures are in good agreement with experiment and previous calculations \cite{ma2009transparent}. However, these static lattice enthalpies do not include zero-point or finite temperature effects, which have been shown to play an important role in the phase stability of Na \cite{gregoryanz2008structural}. Therefore, the thermodynamic stability of candidate phases was studied by calculating their Gibbs free energies within the quasi-harmonic approximation (Section S6).

Comparison of the free energies showed that all of the seven newly discovered structures were preferred over Na $hP$4 at pressure conditions where experiments suggest a rich temperature driven polymorphism (Section S6). However, at the high temperatures consistent with those attained in ramp-compression experiments, the $\mathit{P}6_{3}/\mathit{m}$ phase was computed to possess the lowest Gibbs free energy between 250~GPa (above 710~K) and 350~GPa (above 1900~K). Above 400~GPa, Na $hP$4 again became preferred, in agreement with experimental observations \cite{Denae2022Na}. At 200~GPa (350~GPa) the newly discovered $\mathit{P}6_{3}/\mathit{m}$ structure was computed to be more stable than Na $hP$4 above 150~K (1900~K)  (Figure \ref{fig:latt}(b)). In contrast, the relative enthalpies alone suggest that $\mathit{P}6_{3}/\mathit{m}$ Na is 62~meV/atom \textit{less} stable than $hP$4 Na at 300~GPa, highlighting the importance of the finite temperature effects on the lattice energies.  The absence of imaginary frequencies in the phonon spectra confirmed the dynamic stability of $\mathit{P}6_{3}/\mathit{m}$ Na from 150 to 350~GPa. In addition, first principles molecular dynamics simulations at temperatures typical of those accessed using ramp-compression experiments (1500~K at 260~GPa, and 2000~K at 315~GPa, see Section S7), revealed that the (refined) atomic configurations in $\mathit{P}6_{3}/\mathit{m}$ Na, including the soon-to-be-discussed honeycomb channels, are thermally stable.  At 400~GPa Na $hP$4 possessed the lowest Gibbs free energy to at least 3500~K (Figure S7), in agreement with experimental diffraction data that could be attributed to this phase at 409 $\pm$ 15~GPa \cite{Denae2022Na}.  

\begin{figure}[t!]
\centerline{\includegraphics[width=0.8\columnwidth]{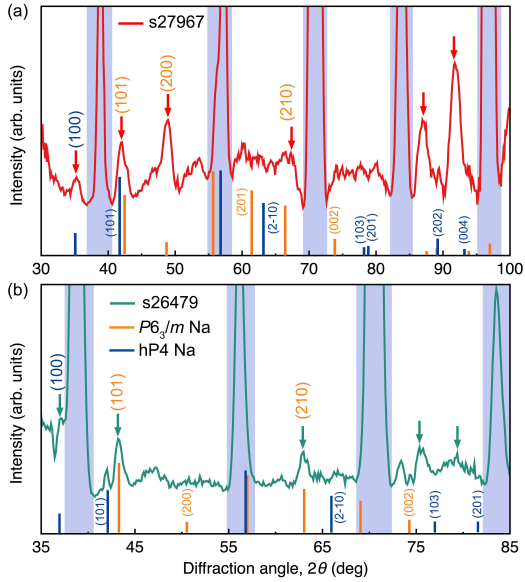}}
\caption{(Color online) Simulated XRD patterns ($\lambda=1.4813$~\AA{}) and experimental lineouts along 2$\theta$ \cite{Denae2022Na} of Na under pressure. The (a) red ((b) green) curves depict the experimental patterns from shot 27697 (26479) at 261 $\pm$ 11~GPa  (315 $\pm$ 11~GPa). The vertical ticks correspond to expected positions and intensities of XRD peaks for the refined $\mathit{P}6_{3}/\mathit{m}$ (yellow) and $hP$4 (blue) symmetry Na structures at 260~GPa (315~GPa). The Na atoms in the $\mathit{P}6_{3}/\mathit{m}$ structure lie on the $6h$ Wyckoff site (0.110, 0.734, 0.250) with refined lattice parameters of a = b = 4.132 \AA{}, and c = 2.472 \AA{} (a = b = 4.015 \AA{}, and c = 2.450 \AA{}). The $hP$4 structure contains Na atoms on the 2a (0, 0, 0) and 2c (1/3, 2/3, 3/4) Wyckoff sites with lattice parameters of a = b = 2.845~\AA{}, and c = 4.062~\AA{} (a = b = 2.705 \AA{}, and c = 4.128 \AA{}, refined). The shaded vertical regions provide the positions of the experimental pinhole calibration peaks.}
\label{fig:xrd}
\end{figure}

To confirm that the predicted $\mathit{P}6_{3}/\mathit{m}$ Na phase was created using laser-driven ramp-compression \cite{Denae2022Na}, we compared the XRD pattern calculated for a refined lattice using the wavelength characteristic of He-$\alpha$ Cu radiation (8.37~keV) with the experimental lineouts along 2$\theta$  at $\sim$260~GPa and $\sim$315~GPa (Figure \ref{fig:xrd}). The $\mathit{P}6_{3}/\mathit{m}$ lattice parameters were refined using a least squares optimization to match its XRD pattern with experimental observations. The refinement considered only the observed diffraction angle $2\theta$'s of the peaks, but not their intensities. Although previously indexed as two different phases ($cI$16, $R\bar{3}m$), the diffraction pattern from $\mathit{P}6_{3}/\mathit{m}$ Na is consistent with the experimental data at both 260 and 315~GPa,  except for the absence of peaks at $\sim$35$^{\circ}$ and $\sim$36$^{\circ}$, respectively. Such low-angle peaks are present in the XRD patterns simulated for Na $hP$4. Due to the small range of static lattice enthalpies between the aforementioned competitive phases (0.3 eV/atom), it is likely that the experimentally detected XRD pattern originates from a mixture of Na phases that coexist due to temperature and pressure gradients within the sample. At high diffraction angles, the 2$\theta$ uncertainty is larger due to transverse pressure gradients that both broaden the diffraction line and contribute a low-pressure tail to the sampled pressure distribution. Therefore, we conclude that the XRD peaks observed in ramp-compressed Na between $\sim$240-325~GPa are consistent with those that would originate from a mixture of the novel $\mathit{P}6_{3}/\mathit{m}$ Na phase, accompanied with Na $hP$4, which  can be caused by the distribution of pressures and temperatures within the relatively thick Na sample layer \cite{Denae2022Na}. Due to preferred orientation, the experimentally observed diffraction intensities cannot be compared to that from an ideal powder.

\begin{figure}[t!]
\centerline{\includegraphics[width=0.9\columnwidth]{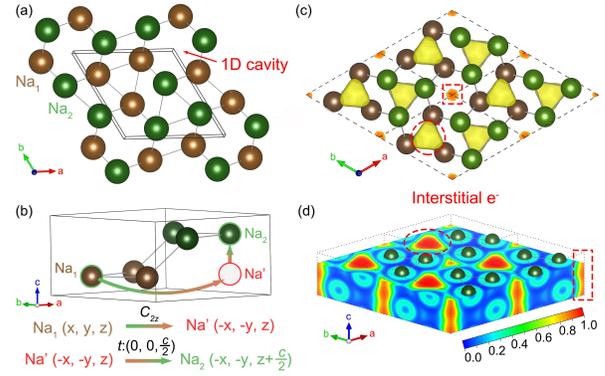}}
\caption{(Color online) (a) The crystal structure of $\mathit{P}6_{3}/\mathit{m}$ Na highlighting the 1D cavity along the $c$-axis by a red arrow. The green and brown atoms indicate the Na atoms lying in two different $ab$ planes. (b) Side view of the unit cell of $\mathit{P}6_{3}/\mathit{m}$ Na. An illustration of a non-symmorphic symmetry operation, $S_{2z}$, that includes a $C_{2z}$ rotation and $\mathit{t} = (0, 0, \frac{c}{2})$ translation is indicated. (c) Valence electron localization function (ELF) with isosurface value of 0.80. (d) Contour map of the ELF, where the red and blue colors refer to the highest (1.0) and the lowest (0.0) values. The dashed red circles (rectangles) in (c) and (d) show that the anionic electrons are located within 0D cages (1D honeycomb channels).}
\label{fig:struc}
\end{figure}

Let us take a closer look at the $\mathit{P}6_{3}/\mathit{m}$ Na phase, which is illustrated in Figure \ref{fig:struc}(a). The structure belongs to a non-symmorphic space group (No.\ 176) lacking mirror, and other roto-inversion symmetries. However, it possesses a non-symmorphic symmetry  (Figure \ref{fig:struc}(b)) generated by a twofold screw rotation along the $c$-axis $S_{2z}$: ($x$, $y$, $z$) $\to$ (-$x$, -$y$, $z$ + $c$/2). This same symmetry operation has been reported to provide protection for topological nodal surface states in crystalline materials \cite{liang2016node,wu2018nodal,wang2019anomalous}. Plots of the ELF (Figure \ref{fig:struc}(c,d)) reveal that $\mathit{P}6_{3}/\mathit{m}$ Na is an electride where paired  electrons are localized in 1D honeycomb channels that run along the $c$-axis, key for the semimetallic nature of this phase, as well as 0D electron blobs centered in Na$_9$ tricapped trigonal prisms. The 1D chains of localized electrons are a striking feature of the ELF plots of both  $\mathit{P}6_{3}/\mathit{m}$ Na and Ba$_3$CrN$_3$ (Figure S20) -- a 1D topological electride characterized by the same spacegroup, which was predicted to possess an anomalous Dirac plasmon with unique properties such as a long lifetime \cite{wang2019anomalous}. 

\begin{figure*}[ht!]
\centerline{\includegraphics[width=0.95\textwidth]{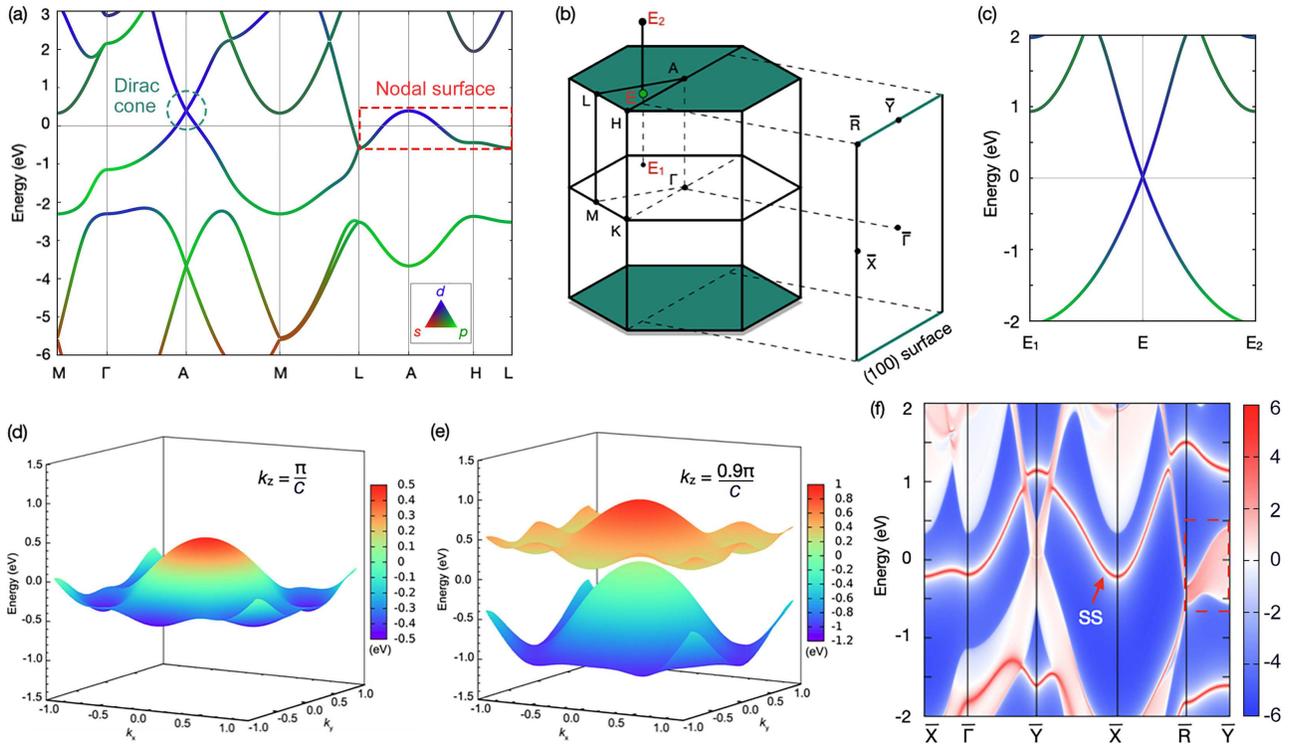}}
\caption{(Color online) (a) Band structure of $\mathit{P}6_{3}/\mathit{m}$ Na at 350 GPa, with the band coloring indicating the orbital character (red/green/blue Na $s$/$p$/$d$). (b) 3D BZ of $\mathit{P}6_{3}/\mathit{m}$ Na and its projection onto a 2D BZ on the (100) surface. High-symmetry and arbitrary ($E_1-E-E_2$) points are labeled. (c) Band structure along the arbitrary $E_{1}-E-E_{2}$ line that lies perpendicular to the $k_{z} = \pi/c$ plane at the arbitrary $E$ point that is located in the plane, as shown in (b). (d,e) The 3D plot of the valence and conduction bands in the $k_{z} = \pi/c$ and $k_z=0.9\pi/c$ plane, respectively. The color represents the energy of the bands relative to $E_F$. The bands are relatively flat and fourfold degenerate in the $k_{z} = \pi/c$ plane (neglecting SOC), which is the nodal surface. The bands split in the $k_{z}$ = 0.9$\pi/c$ plane. (f) Band structure projected on the (100) surface, with the surface state (SS) denoted by a red arrow. The bands enclosed by the red dashed rectangle highlight the projection of the topological nodal surface.}
\label{fig:band}
\end{figure*}

The band structure of $\mathit{P}6_{3}/\mathit{m}$ Na at 350 GPa (Figure \ref{fig:band}(a)) is characteristic of a semimetal. The dispersion is large along the metallic $k_{z}$ direction, confirming the 1D nature of localized electrons within the honeycomb channels. In comparison, the dispersion along the $k_{z}$ = 0 plane with insulating character is relatively small, due to the localized nature of the charge density within this plane. The valence and conduction bands meet in the vicinity of $E_F$ and form a band crossing at the high-symmetry $A$-point. Along the $L-A-H-L$ path, which lies in the $k_{z} = \pi/c$ plane,  the bands stick together and are fourfold degenerate. Plotting the band structure along an arbitrary line, $E_{1}-E-E_{2}$, which is perpendicular to the $k_{z} = \pi/c$ plane at the $E$ point located in this plane (Figure \ref{fig:band}(b)), illustrates that the degeneracy remains at this point. This is seen clearly at $E$ where the Dirac nodal point lies  (Figure \ref{fig:band}(c)),  suggesting that the degeneracy may be present throughout the whole $k_{z} = \pi/c$ plane. Plotting the 3D band structure near $E_F$ while varying $k_x$ and $k_y$, but keeping $k_{z} = \pi/c$ fixed (Figure \ref{fig:band}(d)) confirms our conjecture that the valence and conduction bands are degenerate within this plane. Therefore, the continuous nodal points form a Dirac nodal surface throughout the whole BZ. A plot of the valence and conduction band energies at $k_{z} = 0.9\pi/c$ illustrates that the degeneracy is broken (Figure \ref{fig:band}(e)). Because of the small but non-negligible coupling between the 1D anionic electrons within the honeycomb channels, the Dirac nodal surface is not entirely flat, in-line with previous results for Ba$_{3}$CrN$_{3}$ \cite{wang2019anomalous}. Calculations including spin orbit coupling (SOC) illustrate that the band splitting is negligible for these loosely bound anionic electrons (Figure S14), as expected.

The presence of the Dirac nodal surface in $\mathit{P}6_{3}/\mathit{m}$ is dictated by the non-symmorphic symmetry  $S_{2z}$. The protection mechanism was previously described in Refs.\ \cite{liang2016node,wu2018nodal,wang2019anomalous}, and measured \cite{yang2019observation,xiao2020experimental} in a variety of 3D topological materials. $\mathit{P}6_{3}/\mathit{m}$ Na possesses time-reversal symmetry ($T^{2}$ = 1), and the $TS_{2z}$ operation satisfies $(TS_{2z})^{2}$ = $e^{-ik_{z}}$. As a result, any point on the $k_{z} = \pi/c$ plane is invariant such that $(TS_{2z})^{2}$ = -1, ensuring that a twofold Kramers degeneracy arises at every point on the plane, which is therefore characterized by a topological nodal surface \cite{liang2016node,wu2018nodal}.

The band structure projected on the (100) surface (Figure \ref{fig:band}(f)) illustrates that a surface state band, which does not belong to the projected bulk band structure, crosses $E_F$. Such a metallic surface state is also detected in the band structure projected on the (-110) surface (Figure S13). Meanwhile, the projection of the Dirac nodal surface is observed along the $\overline{R}$ - $\overline{Y}$ path. The presence of metallic surface states illustrates the topological features of the high-pressure, high-temperature $\mathit{P}6_{3}/\mathit{m}$ Na phase.

In summary, \textit{ab initio} evolutionary structure searches coupled with quasiharmonic calculations illustrate that a novel $\mathit{P}6_{3}/\mathit{m}$ symmetry Na phase becomes preferred over the known Na $hP$4 phase at 250~GPa (350~GPa) above 710~K (1900~K).  The XRD patterns obtained in laser-driven ramp-compression experiments at 260 and 315~GPa \cite{Denae2022Na} can be well described by a combination of peaks arising from the $\mathit{P}6_{3}/\mathit{m}$ phase and Na $hP$4. Because of the pressure and temperature gradients in the sample, it is likely that these two phases coexist in experiment. Above 400~GPa Na $hP$4 is predicted to be the ground state structure independent of temperature, in agreement with the experimental observations. 
%Our results highlight the structural complexity of Na close to the melting line, hinting that other phases are waiting to be discovered.  
$\mathit{P}6_{3}/\mathit{m}$ Na is a semimetallic electride owing to 1D chains of interstitial electrons that run along honeycomb channels, and electron blobs localized in Na$_9$ cages. Electronic structure calculations reveal that $\mathit{P}6_{3}/\mathit{m}$ Na possesses a Dirac nodal surface state that is topologically protected by the non-symmorphic $S_{2z}$ symmetry, suggesting it may exhibit unique properties such as anomalous Dirac plasmons predicted to be present in other 1D topological electrides \cite{wang2019anomalous}. We hope this study stimulates further work of Na's structural complexity, topological properties, and core-electron chemistry under extreme conditions of pressure and temperature.

\begin{acknowledgments}
We are grateful to G.W.\ Collins, S.X.\ Hu,  R.J.\ Hemley and S.\ Racioppi for useful discussions.  This material is based upon work supported by the Department of Energy National Nuclear Security Administration under Award Number DE-NA0003856, the University of Rochester, and the New York State Energy Research and Development Authority. K.H.\ acknowledges the Chicago/DOE Alliance Center under Cooperative Agreement Grant No.\ DE-NA0003975.  This material is based upon work supported by the U.S. Department of Energy, Office of Science, Fusion Energy Sciences funding the award entitled \textit{High Energy Density Quantum Matter} under Award Number DE-SC0020340. Partial funding for this research is provided by the Center for Matter at Atomic Pressures (CMAP), a National Science Foundation (NSF) Physics Frontiers Center, under Award PHY-2020249. Computations were carried out at the Center for Computational Research at the University at Buffalo (http://hdl.handle.net/10477/79221).
\end{acknowledgments}
%SM\cite{avery2019xtalopt,kresse1996efficient,kresse1996efficiency,avery2017randspg,lonie2012identifying,blochl1994projector,kresse1999ultrasoft,kresse1996efficient,kresse1996efficiency,perdew1996generalized,monkhorst1976special,baroni2001phonons,gonze1995adiabatic,togo2008first,birch1952elasticity,mostofi2014updated,wu2018wanniertools}

\bibliographystyle{apsrev4-1}
%\bibliography{Ref}
%merlin.mbs apsrev4-1.bst 2010-07-25 4.21a (PWD, AO, DPC) hacked
%Control: key (0)
%Control: author (72) initials jnrlst
%Control: editor formatted (1) identically to author
%Control: production of article title (-1) disabled
%Control: page (0) single
%Control: year (1) truncated
%Control: production of eprint (0) enabled
%

\end{document}